\documentclass[a4paper]{panl}
\usepackage[T2A]{fontenc}
\usepackage[russian,english]{babel}
\usepackage{cite}
\usepackage{wrapfig}
\usepackage{graphicx}
\usepackage{amssymb}
\usepackage{amsfonts}
\usepackage{amsmath}
\usepackage{longtable}
\usepackage{rotating}
\usepackage{lscape}
\usepackage{epsfig}
\usepackage{multirow}

\usepackage{isotope}
\usepackage{tabularx}
\usepackage{booktabs}

\originalTeX
\begin{document}
\title{Detector array for the \isotope[7]{H} nucleus multi-neutron decay study\\
  Массив детекторов для исследования многонейтронного распада ядра \isotope[7]{H}}
\maketitle
\authors{
  \begin{minipage}{\linewidth}
    \centering
    A.~A. Bezbakh$^{a,b}$, S.~G. Belogurov$^a$,
    V. Chudoba$^{a,b}$, A.~S.~Fomichev$^a$,
    A.~V. Gorshkov$^a$,
    L.~V.~Grigorenko$^a$, G.~Kaminski$^{a,c}$, M.~S.~Khirk$^a$
    A.~G.~Knyazev$^a$,
    S.~A.~Krupko$^a$,
    B.~Mauyey$^{a,d}$,
    I.~A.~Muzalevskii$^{a,b}$, E.~Yu.~Nikolskii$^a$,
    A.~M.~Quynh$^{a,e}$,   P.~G.~Sharov$^{a,b}$,
    R.~S.~Slepnev$^a$,  S.~V.~Stepantsov$^a$,
    G.~M.~Ter-Akopian$^a$,
    and R.~Wolski$^a$
  \end{minipage}
}
 
\setcounter{footnote}{0}
\authors{
  \begin{minipage}{\linewidth}
    \centering
    А.~А.~Безбах$^{a,b}$, С.~Г.~Белогуров$^a$, В.~Худоба$^{a,b}$,
    А.~С.~Фомичев$^a$, А.~В.~Горшков$^a$, 
    Л.~В.~Григоренко$^a$, Г.~Каминьски$^{a,c}$, М.~С.~Хирк$^a$, А.~Г.~Князев$^a$, С.~А.~Крупко$^a$,
    Б.~Мауей$^{a,d}$, И.~А.~Музалевский$^{a,b}$, Е.~Ю.~Никольский$^a$,
    А.~М.~Куинь$^{a,e}$, П.~Г.~Шаров$^{a,b}$,
    Р.~С.~Слепнев$^a$, 
    С.~В.~Степанцов$^a$, Г.~М.~Тер-Акопьян$^a$, Р.~Вольски$^a$
\end{minipage}
}
\from{$^a$\,Flerov Laboratory of Nuclear Reactions, JINR, 141980 Dubna, Russia\\
  $^a$\,Лаборатория ядерных реакций им. Г.Н. Флерова, ОИЯИ, 141980 Дубна, Россия
}
\from{$^b$\,Institute of Physics, Silesian University in Opava, 74601 Opava, Czech Republic\\
  $^b$\,Физический институт, Силезский университет в Опаве, 74601 Опава, Чехия}
\from{$^c$\,Heavy Ion Laboratory, University of Warsaw, 02-093 Warsaw, Poland\\
  $^c$\,Лаборатория тяжелых ионов, Варшавский университет, 02-093 Варшава, Польша}
\from{$^d$\,L.N. Gumilyov Eurasian National University, 010008 Astana, Kazakhstan\\
  $^d$\,Евразийский национальный университет имени Л. Н. Гумилёва, 010008 Астана, Казахстан}
\from{$^e$\,Nuclear Research Institute, 670000 Dalat, Vietnam\\
  $^e$\,Институт ядерных исследований, 670000 Далат, Вьетнам}

\begin{abstract}
 \begin{otherlanguage}{russian}

    На пучке радиоактивных ядер сепаратора ACCULINNA-2
    Лаборатории ядерных реакций  им. Г. Н. Флёрова создаётся установка,
    предназначенная для детального изучения
    пятительного распада ядер \isotope[7]{Н}, образующихся в результате реакции
    передачи протона от налетающего ядра \isotope[8]{He} на ядро мишени
    \isotope[2]{Н}.
    В настоящей статье даётся описание устройства из 100 пластических
    сцинтилляторов ВС-404, предназначенного для регистрации нейтронов,
    телескопа Si детекторов ядер отдачи \isotope[3]{He}, а также комплекта
    детекторов для \(\Delta E\)-\(E\)-TOF регистрации ядер \isotope[3]{H},
    вылетающих из мишени при распаде \isotope[7]{H}.
    Показаны результаты моделирования методом Монте-Карло, давшего ожидаемые
    значения энергии и параметры траекторий всех частиц.
    Приведены результаты полученных оценок светимости обсуждаемых экспериментов.
  \end{otherlanguage}
  
  \vspace{0.2cm}
  
  Setup fitting the requirements for the detailed study of the five-body decay
  of the \isotope[7]{H} nucleus obtained as a result of the proton transfer from
  the \isotope[8]{He} projectiles to the deuterium target nuclei is being built
  at the radioactive beam line of ACCULINNA-2 separator in the
  G.N. Flerov Laboratory of Nuclear Reactions.
  Described here is the assembly of 100 BC-404 plastic scintillators,
  intended for neutron detection, the annular Si detector telescope for the
  isotope[3]{He} recoils, and the detector array providing the
  \(\Delta E\)-\(E\)-TOF registration of \isotope[3]{H} nuclei emitted at the
  \isotope[7]{H} decay.
  Results obtained by the Monte Carlo simulations made for the energy values and
  flight passes of all these particles are given together with the luminosity
  expected for the discussed experiments.

\end{abstract}
\vspace*{6pt}

\vspace*{6pt}

Submitted to PEPAN letters.

\section{Setup with the 75 mm thick BC-404 scintillators}

\begin{figure}
  \centering
  \includegraphics[width=\linewidth]{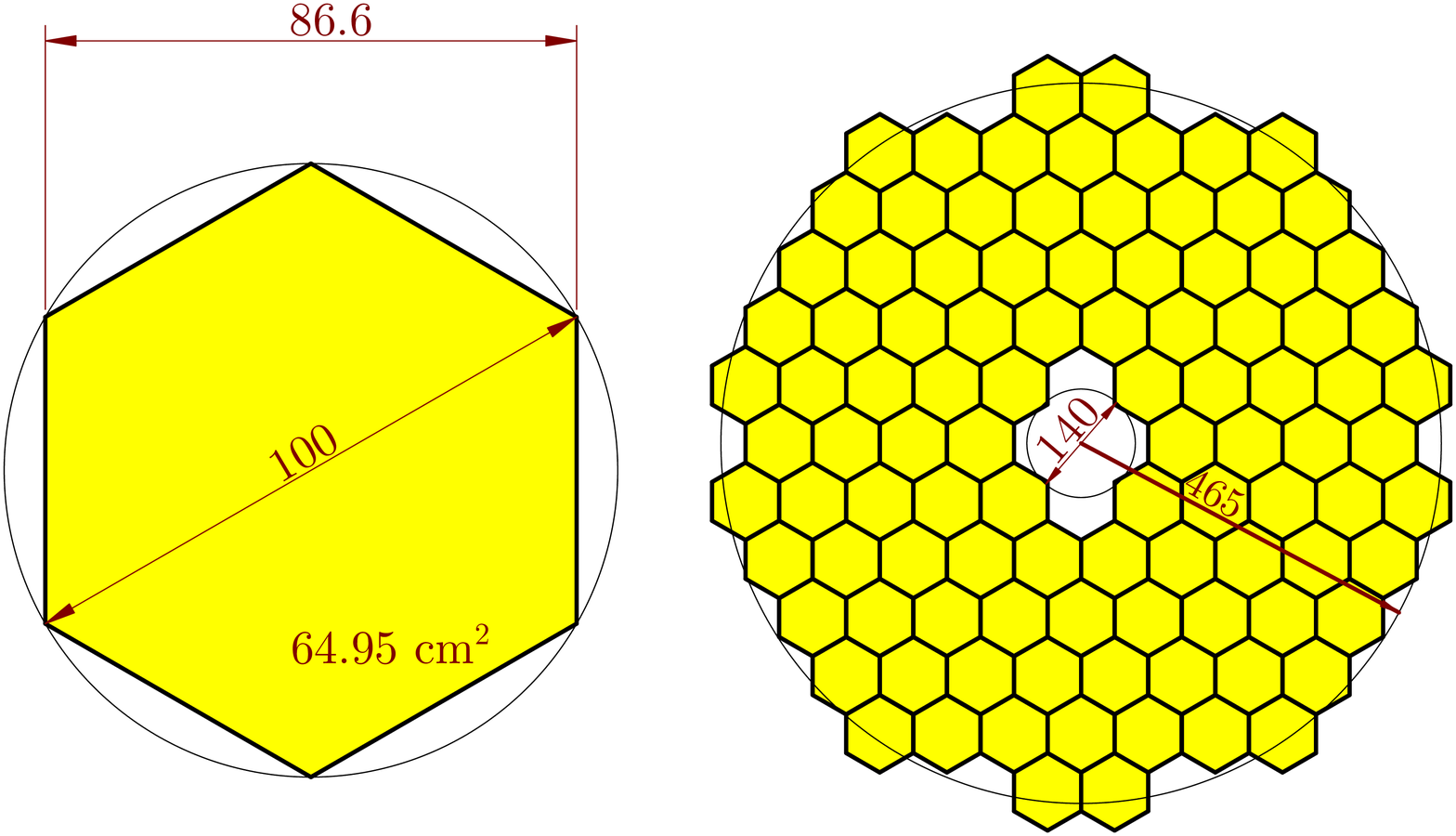}
  \caption{\label{fig:plastic}
    Left: The 75-mm thick plastic scintillator having the hexagonal
    cross-section.
    Painted in yellow is shown the plastic cross-section
    enclosed in the lightproof (aluminum) shield.
    Right: Assembly of 100 BC-404 plastic scintillators suggested for
    the detection of nuclear decay events with multiple neutron emission.}
\end{figure}

Presented is the modification of the detector array intended for the study of
nuclei undergoing decay with the multi-neutron emission.
The modification assumes the use of the assembly of 100 BC-404 scintillators
which are at the disposal of ACCULINNA group.
The front view of a single scintillator is shown in Figure~\ref{fig:plastic}.
In its cross-section, the 75 mm thick plastic is the hexagon inscribed in
the circle with diameter 100 mm and it is enclosed in some 1-mm thick cover.
Consequently, 95.5\% area of a compact assembly made with the use of these
plastics will be covered by the BC-404 scintillators.

\begin{figure}
  \centering
  \includegraphics[width=0.75\linewidth]{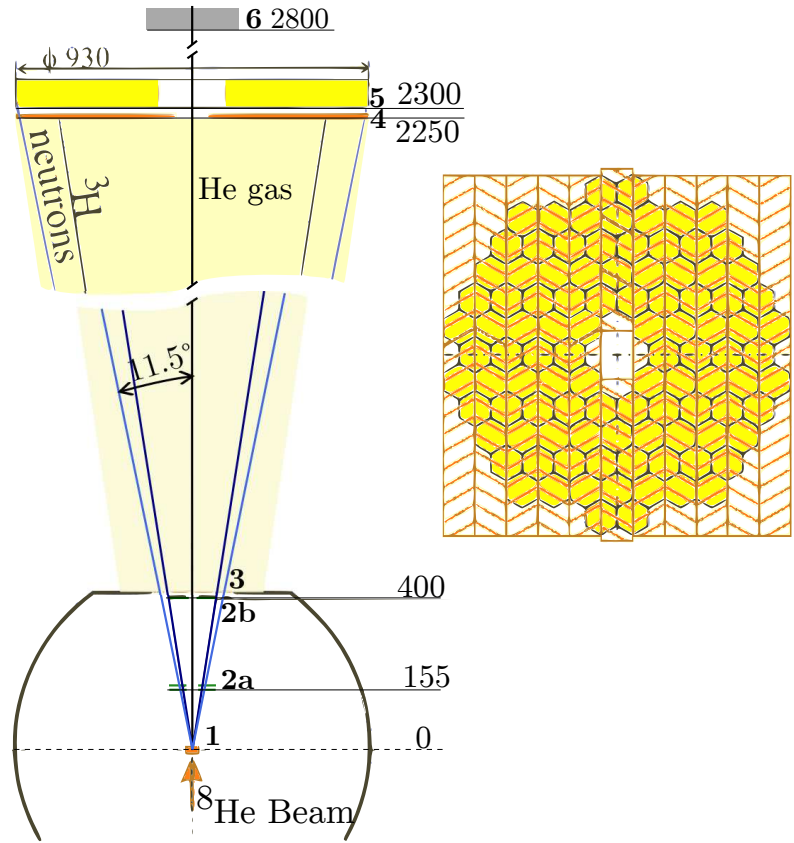}
  \caption{\label{fig:setup}
    Setup for the study of the \isotope[7]{H} decay events:
    1 --- \isotope[2]{H} cryogenic deuterium target;
q    2a --- 0.140 mm thick annular Si-detector backed with the
    1 mm thick annular Si detector;
    2b --- 1.5 mm thick annular Si-detector;
    3 --- 0.18 mm thick stainless-steel vacuum window of
    the ACCULINNA-2 reaction chamber;
    4 --- plastic-scintillator array stopping the \isotope[7]{H}-decay tritons;
    5 – Neutron Wall (NW);
    6 – beam stopper.
    Neutrons emitted from the target within angular range
    \(\theta_{\text{lab}}\le 11.5^\circ\) at the \isotope[7]{H} decay hit the NW.
    The distances from the target to the items enumerated in this Figure
    are given in millimeters on the right side.
    Shown on the right, in insert, is the front view of array 4
    with the NW installed behind.}
\end{figure}

The Neutron Wall (NW) arrangement, shown in Figure~\ref{fig:plastic},
will be the central part of the setup being created to satisfy the requirements
becoming apparent as a result of study made on the decay spectra of
\isotope[7]{H} and \isotope[6]{H}
nuclei~\cite{Bezbakh:2020,Muzalevskii:2021,Nikolskii:2022}.
The setup composition is given in Figure~\ref{fig:setup}.

Assembly of 100 plastic scintillators shown in Figure~\ref{fig:plastic} covers
a surface approximated by a circle with diameter 93 cm.
Its empty central part opens way for the \isotope[8]{He} nuclei to fly free
to the beam stopper placed at a 0.5-meter distance behind the NW without making
background distorting the detection of multiple-neutron events.

Detection probability\(\varepsilon_n\approx 0.28\) is estimated for
single neutrons with energy 20--30 MeV hitting the array.
This estimate is based on the known~\cite{Kohley:2012,Tippawan:2009}
cross section values of the elastic \(\isotope[1]{H}(n,p)\) scattering and
\(\isotope[12]{C}(n,n'3\alpha)\), \(\isotope[12]{C}(n,\alpha)\isotope[9]{Be}\),
\(\isotope[12]{C}(n,np)\isotope[11]{B}\),
\(\isotope[12]{C}(n,p)\isotope[12]{B}\), \(\isotope[12]{C}(n,n\gamma)\)
reactions induced by neutrons bombarding the hydrogen and carbon nuclei making
the BC-404 plastic composition in amounts \(5.23\times 10^{22}\) cm\(^{-3}\) and
\(4.74\times 10^{22}\) cm\(^{-3}\), respectively.

The depletion rate of neutron flux in the BC-404 plastic is calculated by
the use of the depletion constant \(\lambda\).
The value of this constant is obtained as the sum of contributions made to
the decrease of the neutron flux due to the neutron elastic scattering on
the hydrogen nuclei and neutron reactions with carbon nuclei giving rise to
the charged-particle emission.
The values of this depletion constant obtained for neutrons with five different
energy values are given in Table~\ref{tab:depl-const}.

\begin{table}
  \caption{\label{tab:depl-const}
    Depletion constant and flux-decrease factors in the material of
    plastic BC-404 obtained for the neutrons at five energy values.
    \(F\) is  factor of flux reduction at plastic depth 75 mm.}
  \begin{tabularx}{\linewidth}{lXrrrrr}
    \toprule
    \(E_n\), MeV&&   10&   20&   25&   30&   35\\
    \midrule
    \(\lambda\), mm\(^{-1}\)&& 0.00760& 0.00770& 0.00738& 0.00724& 0.00630\\
    \(F\)&&    0.56&    0.56&    0.57&    0.58&    0.68\\
    \bottomrule
  \end{tabularx}
\end{table}

The values of flux reduction factor obtained as \(F = e^{-\lambda\cdot 75}\) for
neutrons passing through the 75-mm thick BC-404 scintillator are given in
Table~\ref{tab:depl-const} in last raw.
Thus, the probability that neutrons with energy 20--30 MeV could produce any of
the specified above reaction is estimated as \(1-F\approx 0.43\).
Detection probability\(\varepsilon_n approx 0.28\) was obtained comparing these
reaction probabilities with those which were typical for the neutron modular
detector DEMON~\cite{Tilquin:1995}.

\section{Triggering array}

The suggested choice for the detector array, giving trigger signals produced by
the \isotope[3]{He} recoil nuclei emitted in the
\(\isotope[2]{H}(\isotope[8]{He},\isotope[3]{He})\isotope[7]{H}\) reaction,
is to place at position 2a (see Figure~\ref{fig:setup}) a detector telescope
made of a pair of annular Si detectors.
Each detector is segmented in 64 rings and 64 sectors and has the 28-mm central
hole and sensitive area with the 32 mm inner diameter and 125 mm outer diameter.
The 0.140 mm thick front detector is the source of trigger signals generated
from the recoil \isotope[3]{He} nuclei.
The second detector will be 1.0 mm thick.

Energy range 9--22 MeV is inherent to the \isotope[3]{He} recoils emitted in the
\(\isotope[2]{H} (\isotope[8]{He},\isotope[3]{He}) \isotope[7]{H}\) reaction\
populating the \isotope[7]{H} nucleus excitation spectrum in a range of 0--9 MeV
above the \(\isotope[3]{H}+4n\) decay threshold. Such recoils with energy
\(> 14\) MeV pass through the 0.140 mm front Si detector and are stopped in the
second, 1~mm thick Si detector.
These nuclei will be identified by the \(\Delta E-E\) method against the
background made by the other recoil nuclei
(\isotope[2]{H}, \isotope[3]{H}, \isotope[4]{He}, \isotope[6]{H}e, etc.)
which are emitted from the target.
The 9--14 MeV \isotope[3]{He} recoils, stopped in the front Si detector,
will be safely discriminated from the \isotope[3]{H} recoils, emitted from
the target with similar energy, because for these \isotope[3]{H} recoil nuclei
the \(\Delta E-E\) identification signals are provided by the detector telescope
placed in position 2a.

As for the \isotope[3]{He} and \isotope[4]{He} recoil nuclei emitted from the
target with energy \(E_{\isotope[3]{He}}< 14\) MeV and
\(E_{\isotope[4]{He}}< 16\) MeV, respectively, their identification will be
performed on the basis of the \(E\)-ToF method consisting in the comparison of
the measured energy deposit made by the recoil nucleus in the front, 0.140 mm
detector, and its flight time (ToF) expended on the 155 mm distance from
the target to this detector.
The ``start'' (zero-mark) time is given by the radioactive-beam (RIB) diagnostic
array situated upstream the target, and the signal coming from the front,
0.14 mm \(\Delta E\) detector will provide the stop mark for the ToF evaluation.
The zero-time start signal is provided by the ACCULINNA-2 RIB diagnostic with a
root-mean-square error 0.2 ns, and the time measurement made with the
\(\Delta E\) Si detector gives the 0.5 ns stop-signal time spread.
This will result in the safe separation of individual nuclides between
the whole group of the \isotope[3]{He}, \isotope[4]{He}, \isotope[6]{H}e,
\isotope[8]{He}, \isotope[6]{Li}, and \isotope[7]{Li} recoil nuclei hitting
the 0.14 mm Si detector.
Results obtained by means of Monte Carlo (MC) simulations testify in behalf of
this estimation.
All simulations assumed that the \isotope[8]{He} beam, having emittance
\(75~\pi~\text{mm}\cdot\text{mrad}\) and energy spread
\(\Delta E/E=5\%\) (FWHM), was focused on a target within a circle with
diameter 15 mm.

\begin{figure}
  \centering
  \includegraphics[width=0.84\linewidth]{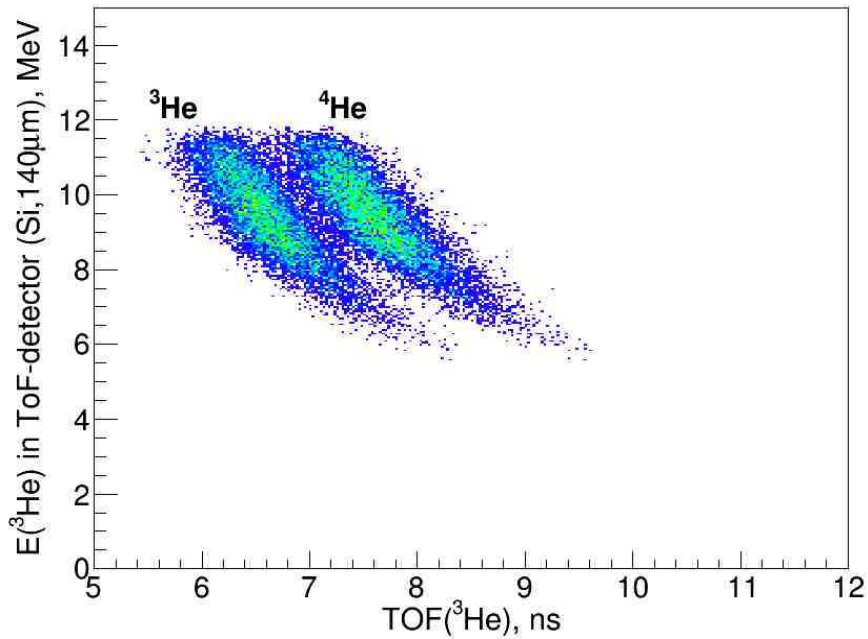}
  \caption{\label{fig:tof-spectra}
    The ToF spectra simulated for the \isotope[3]{He} and \isotope[4]{He} nuclei
    emitted from the target and stopped in the 0.140 mm annular Si detector.
    Further explanations are given in text.}
\end{figure}

One simulation result is presented in Figure~\ref{fig:tof-spectra}.
Shown are the Energy -- ToF distributions simulated for the \isotope[3]{He}
recoil nuclei and \isotope[4]{He} background nuclei emitted from the target and
detected by the 0.140 mm thick annular Si detector 2a shown in
Figure~\ref{fig:setup}.
The source of the \isotope[3]{He} recoil nuclei is the
\(\isotope[2]{H}(\isotope[8]{He},\isotope[3]{He})\isotope[7]{H}\) reaction
populating the \isotope[7]{H} 2.2 MeV ground state in the center-of-mass angular
range 0--20 degrees.
These nuclei are assumed to be formed in the thick, 2.5 mg/cm\(^2\) cryogenic
deuterium target.
It is evident that the distributions, simulated for the two recoils, only weakly
overlap in this two-dimension plot.
More than 95\% of these \isotope[3]{He} recoil nuclei can be separated without a
noticeable \isotope[4]{He} impurity.
Being taken in 1.0-MeV \isotope[3]{He} energy bands, these distributions have
roughly 0.5 ns width (FWHM).

\section{Information deduced on the \isotope[7]{H} spectrum
  using the combined mass method}

We will discuss now the application of the combined-mass method,
originally developed in~\cite{Sharov:2017}, to the study of \isotope[7]{H}
spectrum obtained in the
\(\isotope[2]{H}(\isotope[8]{He},\isotope[3]{He})\isotope[7]{H}\) reaction.
Discussed here is the experiment which will be carried out with the
25 \(A\)\*MeV radioactive \isotope[8]{He} beam bombarding a 2.5 mg/cm\(^2\)
(\(7.5\times 10^{20} ~\text{cm}^{-2}\)) deuterium gas target.
So thick target leads to the \isotope[7]{H} missing mass determination
practically useless with very bad energy resolution.
However, the poorly measured energy and emission angle of the \isotope[3]{He}
recoil nucleus cause the good determination of the center-of-mass momentum value
and the trajectory angle made for complete pattern of five decay products
(\(\isotope[3]{H}+4n\)) emitted by the \isotope[7]{H} nucleus in
the \(\isotope[2]{H}(\isotope[8]{He},\isotope[3]{He})\isotope[7]{H}\) reaction.
Precision achieved by this procedure depends on the errors inherent to
the measured \isotope[3]{He} energy (\(E_{\text{lab}}\)) and angle
(\(\theta_{\text{lab}}\)).

\begin{figure}
  \centering
  \includegraphics[width=0.84\linewidth]{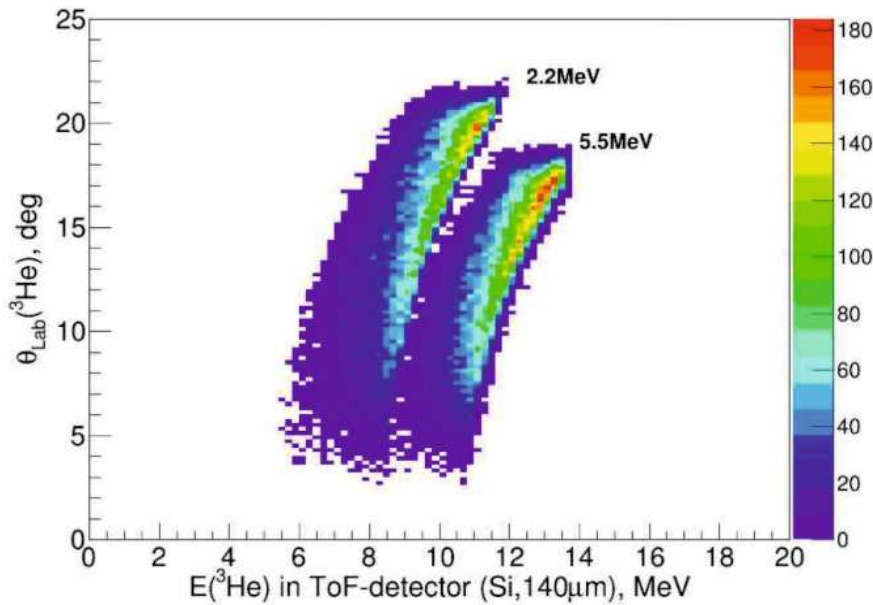}
  \caption{\label{fig:theta-E}
    The \(\theta_{\text{lab}}\) vs. \(E_{\text{lab}}\) distributions obtained
    by the complete MC simulations made for \isotope[3]{He} recoil nuclei
    appearing in the
    \(\isotope[2]{H}(\isotope[8]{He},\isotope[3]{He})\isotope[7]{H}\)
    reaction resulting in the population of the \isotope[7]{H} resonance states
    with energy 2.2 and 5.5 MeV.
    Simulations were made for the cryogenic D\(_2\) target with a thickness of
    \(2.5~\text{mg}\cdot\text{cm}^{-2}\), in the center-of-mass angular range
    0--20 degrees.}
\end{figure}

Realistic estimates come out for these errors from the complete MC simulation
made for the \(\isotope[2]{H}(\isotope[8]{He},\isotope[3]{He})\isotope[7]{H}\)
reaction populating the known \isotope[7]{H} resonance
states~\cite{Muzalevskii:2021} with energies 2.2 and 5.5 MeV above
the \(\isotope[3]{H}+4n\) decay threshold.
The \(\theta_{\text{lab}}\) vs. \(E_{\text{lab}}\) distributions obtained for
\isotope[3]{He} recoils showing the population of these \isotope[7]{H} states
are presented in Figure~\ref{fig:theta-E}.
The pattern displayed in Figure~\ref{fig:theta-E} shows that less than 1.5\% of
5.5-MeV states populated in the
\(\isotope[2]{H}(\isotope[8]{He},\isotope[3]{He})\isotope[7]{H}\) reaction mix
with the 2.2-MeV states.
With \(\theta_{\text{lab}}\) measured with a 20 mrad precision, one knows,
with precision better than 4 mrad, the center-of-mass emission angle of
the produced \isotope[7]{H} nucleus.
At that time, the\isotope[7]{H} momentum value in laboratory system is defined
with accuracy 0.05\%.
For that reason, the accurate knowledge of the\isotope[7]{H} momentum vector
together with information about only four of the five
\(\isotope[7]{H}\to\isotope[3]{H} + 4n\) decay products enables one to
reconstruct the \isotope[7]{H} excitation energy with high resolution.

The discussed approach will allow one to see the real profiles of the
ground state (2.2 MeV) and the first excited state (5.5 MeV) \isotope[7]{H}
resonances with the FWHM resolution making 0.45 MeV and 0.75 MeV, respectively.
These values are almost 2 times better than the resolutions achieved
in~\cite{Bezbakh:2020,Muzalevskii:2021}.

\section{\isotope[3]{H} nuclei and neutrons emitted at the \isotope[7]{H} decay}

The detection of \isotope[7]{H} decay neutrons, their time-of flight and
trajectory measurement, is the principal function of
the Neutron Wall presented above.
Operating together with the RIB diagnostic system of the ACCULINNA-2 separator
the Neutron Wall placed at a distance 230 cm downstream the target will provide
one-percent precision in the velocity measurement done for the neutrons.
Their trajectory angles will be measured with accuracy \(\pm 20\) mrad.

The 1.5-mm thick annular Si multi-strip detector 2b
(see Figure~\ref{fig:setup}), having the same sensitive area as the recoil
detectors 2a, will give \(\Delta E\) signal for the main part of the
\isotope[3]{H} nuclei emitted from the target offering the measurement of their
specific energy loss (\(\Delta E\)) and emission angle (\(\theta_{\text{lab}}\)
made with accuracy 150 keV and 7.5 mrad, respectively.
Some of these \isotope[3]{H} nuclei, emitted at the \isotope[7]{H} decay at
larger lab angles, will miss 2b.
But information on their specific energy loss and emission angles will come from
the \(\Delta E-E\) array 2a.

The plastic detector array 4 is placed in front of NW
(see this Figure~\ref{fig:setup}) denoted in the whole setup and shown in
front view in insert given in the Figure~\ref{fig:setup} right side.
It is intended for measuring the rest energy \(E\) and the flight time (ToF) of
\isotope[3]{H} nuclei at the distance from the target making 225 cm.
It is made of ten 1-meter long and 88-mm wide BC-404 scintillator plates having
thickness equal 20 mm.
Being disposed as two groups in five plates these plastics will be placed on
the two sides of axis Y, marked in Figure~\ref{fig:plastic}.
Each scintillator plate is viewed from two sides by the PMTs resulting in
the definition of \isotope[3]{H} stop time made with 0.2-ns accuracy.
The \isotope[3]{H} nuclei emitted from the target with energy \(\le 100\) MeV
are stopped in the 20 mm thick plastic.

The pair of 45 cm long, 20-mm thick plastics, each coupled with PMT on one side,
will be placed in front of NW to cover its middle area left uncovered by the
1-meter plastics.
This plastic pair will leave open free way for \isotope[8]{He} nuclei to pass to
the beam stopper.
Thus, the ToF-\(\Delta E\)-\(E\) identification will be provided for
the \isotope[3]{H} nuclei coming from the target.
The ToF values measured with resolution 0.4 ns will give energy determination
made for the \isotope[3]{H} nuclei emitted at the five-body decay of
\isotope[7]{H} nucleus.

\begin{figure}
  \centering
  \includegraphics[width=0.84\linewidth]{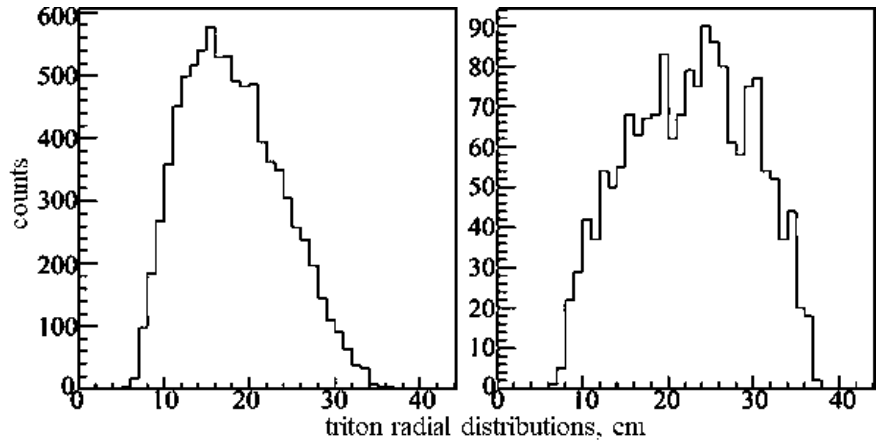}
  \caption{\label{fig:radial-distribution}
    MC simulated radial distributions obtained at the distance 225 cm from
    the target for the \isotope[7]{H} decay tritons.
    On the left and right sides shown are the distributions obtained for
    the decay of the ground (2.2 MeV) and first (5.5 MeV) \isotope[7]{H} states,
    respectively, populated in the
    \(\isotope[2]{H}(\isotope[8]{He},\isotope[3]{He})\isotope[7]{H}\)
    reaction in the center-of-mass angular range 0 – 20 degrees.}
\end{figure}

\begin{figure}
  \centering
  \includegraphics[width=0.84\linewidth]{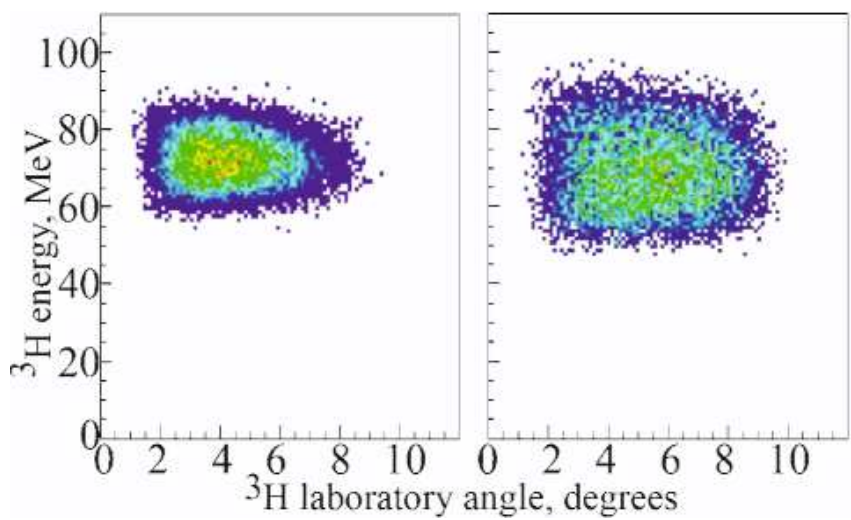}
  \caption{\label{fig:theta-E-3H}
    Energy vs. angle distributions obtained by MC simulation for
    the \isotope[3]{H} nuclei emitted at the 5-body decay of \isotope[7]{H}
    nuclei produced in the
    \(\isotope[2]{H}(\isotope[8]{He},\isotope[3]{He})\isotope[7]{H}\) reaction
    in their 2.2 MeV ground state (left side) and in the 5.5 MeV first resonance
    state (right side).}
\end{figure}

MC simulations were made for the \isotope[7]{H} five-body decay in the framework
of phase-space volume approximation.
In particular, shown in Figure~\ref{fig:radial-distribution} are
the MC simulated radial distributions obtained at the distance 225 cm from
the target for the \isotope[7]{H} decay tritons.
Patterns in Fig.~\ref{fig:theta-E-3H} allow one to see the energy values of
the \isotope[3]{H} nuclei shown against their emission angles.
Presented in this figure are the simulation results where the detection of
the \isotope[3]{H} recoils, emitted at the reaction center-of-mass angles
0--20 degrees, is supplemented with the arrival of all \isotope[7]{H}
decay products, \isotope[3]{H} and 4 neutrons, to the plastic array 4 and NW 5,
respectively (see Figure~\ref{fig:setup}).

\begin{figure}
  \centering
  \includegraphics[width=0.84\linewidth]{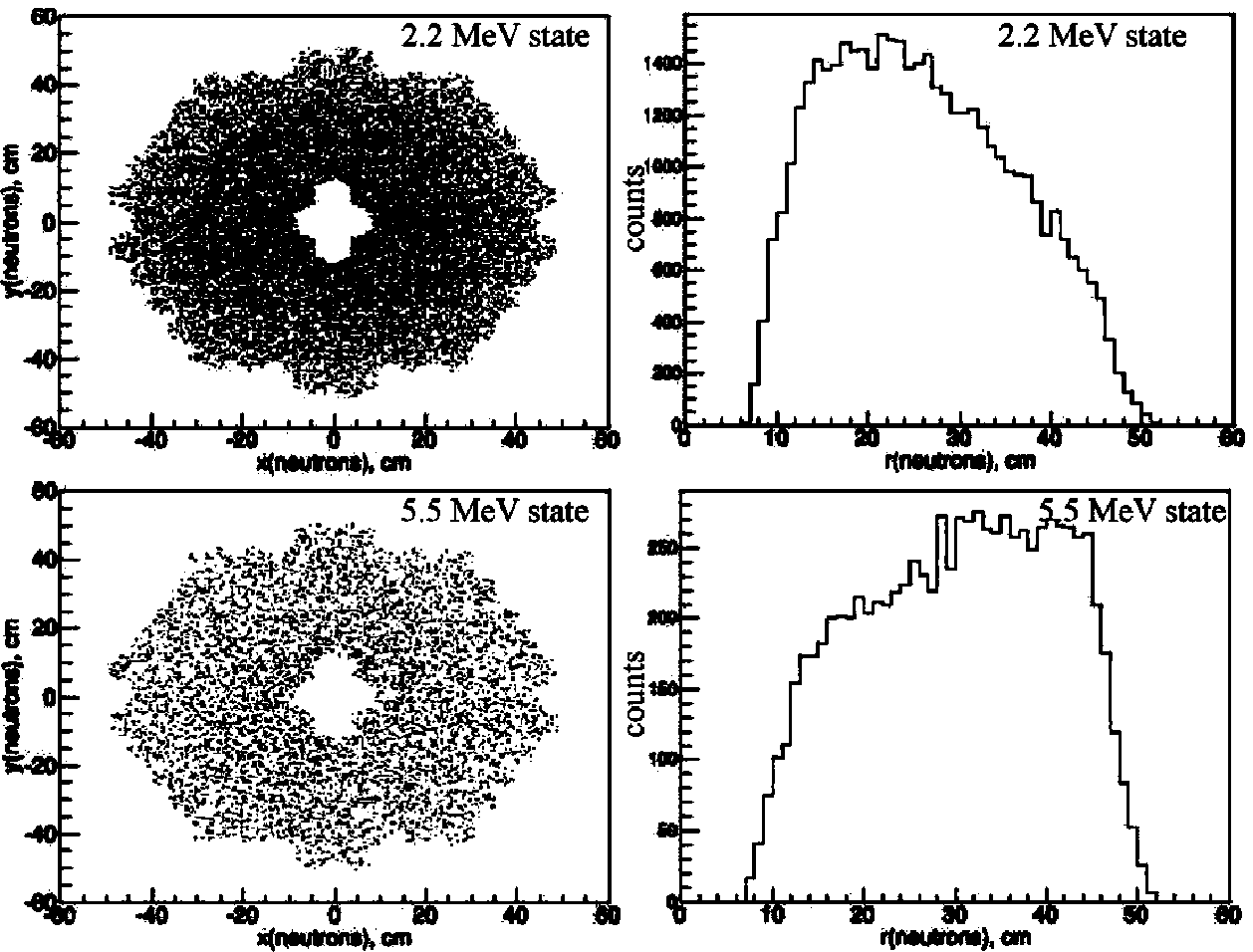}
  \caption{\label{fig:profiles}
    MC simulated transverse profiles (left column) and radial distributions
    (right column) obtained at the Neutron Wall for the \isotope[7]{H} decay
    neutrons emitted from the ground (2.2 MeV) and the first excited (5.5 MeV)
    \isotope[7]{H} states.}
\end{figure}

\begin{figure}
  \centering
  \includegraphics[width=0.84\linewidth]{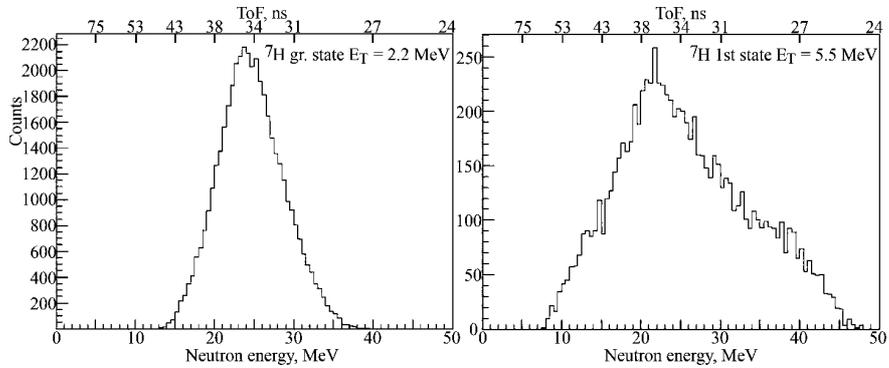}
  \caption{\label{fig:n-spectra}
    The energy spectra of neutrons emitted from the deuterium target when
    \isotope[7]{H} is produced in its ground (2.2 MeV) and the first excited
    (5.5 MeV) states in the
    \(\isotope[2]{H}(\isotope[8]{He},\isotope[3]{He})\isotope[7]{H}\) reaction.
    Given on top are the neutron ToF values calculated for the 230 cm distance
    to the Neutron Wall.}
\end{figure}

Distributions simulated for the neutron arrival positions at the Neutron Wall
are shown in Figure~\ref{fig:profiles}.
Looking at the neutron energy spectra shown in Figure~\ref{fig:n-spectra}
one can see a quite large spread in the energy and ToF values.

\section{Luminosity estimation}

Basic for the estimations made on the setup efficiency is the chance that
minimum four of the five \isotope[7]{H} decay products arrive at
the detectors 2a, 2b, 4 and 5 (see Figure~\ref{fig:setup}) when the
\isotope[3]{He} recoil emitted in the
\(\isotope[2]{H}(\isotope[8]{He},\isotope[3]{He})\isotope[7]{H}\) reaction
is detected.
The results of such estimations made on the basis of MC simulation are given
in Table~\ref{tab:prob-geom}.
\begin{table}
  \caption{\label{tab:prob-geom}
    Probabilities to have the \isotope[7]{H} decay products arriving at
    the position in front of the Neutron Wall in coincidence with
    the \isotope[3]{He} recoil detection.}
  \begin{tabularx}{\linewidth}{Xcccccc}
    \toprule
    \isotope[7]{H} states&
    \multicolumn{2}{c}{g. s., 2.2 MeV}&
    \multicolumn{2}{c}{1st state, 5.5 MeV}&
    \multicolumn{2}{c}{2nd state, 7.5 MeV}\\
    \(\theta_{\text{cm}}\) (deg)& 0--20&20--30& 0--20&20--30& 0--20&20--30\\
    \midrule
    \(\isotope[3]{He}+t4n\)&0.25&0.27&0.049&0.032&0.019&0.012\\
    \(\isotope[3]{He}+t3n\)&0.32&0.36&0.29&0.22&0.18&0.14\\
    \(\isotope[3]{He}+ 4n\)&0.065&0.027&0.013&0.019&0.008&0.009\\
    \bottomrule
  \end{tabularx}
\end{table}

Knowing these results and taking from Table~\ref{tab:depl-const} the values o
f neutron flux decrease and depletion constant \(\lambda\), the detection
probabilities were estimated for the particle groups indicated in
Table~\ref{tab:prob-geom}.
These results are given in Table~\ref{tab:prob-reg}.

\begin{table}
  \caption{\label{tab:prob-reg}
    Probabilities (\(\varepsilon_d\)) estimated for the
    \isotope[3]{He}-triggered detection of minimum four particles emitted at
    the five-body \isotope[7]{H} decay.}
  \begin{tabularx}{\linewidth}{Xccccccc}
    \toprule
    \isotope[7]{H} states&&
    \multicolumn{2}{c}{g. s., 2.2 MeV}&
    \multicolumn{2}{c}{1st state, 5.5 MeV}&
    \multicolumn{2}{c}{2nd state, 7.5 MeV}\\
    \(\theta_{\text{cm}}\) (deg)&& 0--20&20--30& 0--20&20--30& 0--20&20--30\\
    \midrule
    \(\isotope[3]{He}+t4n\)&\(t4n\)&0.0015&0.0017&0.0003&0.0002&0.0001&0.0001\\
                         &\(t3n\)& 0.022& 0.024&0.0043&0.0028&0.0017&0.0011\\
    \(\isotope[3]{He}+t3n\)&&0.0070&0.0079&0.0064&0.0048&0.0040&0.0031\\
    \(\isotope[3]{He}+ 4n\)&\(4n\)&0.0004&0.0002&0.0001&0.0001&0.0000&0.0001\\
    \(\isotope[3]{He}+ 4n\)&\(3n\)&0.0057&0.0024&0.0011& 0.0017&&\\
    \(\varepsilon_d\), all together&&0.037&0.0362&0.0122&0.0096&0.0065&0.0052\\
    \bottomrule
  \end{tabularx}
\end{table}

Detection probabilities presented in Table~\ref{tab:prob-reg} (last row) offer
the way to calculate the luminosity achievable in the experiment planned for the
study of \isotope[7]{H} energy spectrum populated in the transfer reaction
\(\isotope[2]{H}(\isotope[8]{He},\isotope[3]{He})\isotope[7]{H}\).
Predictions made for the luminosity \(L\) appearing for the \isotope[7]{H}
formed in the three energy states, 2.2, 5.5, and 7.5 MeV are presented
in Table~\ref{tab:lum-est}.
\begin{table}
  \caption{\label{tab:lum-est} Luminosity estimations.}
  \begin{tabularx}{\linewidth}{Xcccccc}
    \toprule
    \isotope[7]{H} states
    &\multicolumn{2}{c}{g. s., 2.2 MeV}
    &\multicolumn{2}{c}{1st state, 5.5 MeV}
    &\multicolumn{2}{c}{2nd state, 7.5 MeV}\\
    \(\theta_{\text{cm}}\) (deg)& 0--20&20--30& 0--20&20--30& 0--20&20--30\\
    \midrule
    \(\varepsilon_d\) &0.037&0.0362&0.0122&0.0096&0.0065&0.0052\\
    \(\Delta\Omega\), sr&0.35&0.08&0.33&0.27&0.31&0.46\\
    \(L,~\text{cm}^{-2}\text{s}^{-1}\)
    &\(8.5\!\cdot\! 10^{23}\)
    &\(2.2\!\cdot\! 10^{23}\)
    &\(3.0\!\cdot\! 10^{23}\)
                                &\(2.0\!\cdot\! 10^{23}\)
                                       &\(1.5\!\cdot\! 10^{23}\)
                                              &\(1.8\!\cdot\! 10^{23}\)   \\
    \bottomrule
  \end{tabularx}
\end{table}

Luminosity is calculated as the product
\[
  L=\varepsilon_d\times t\times I \times \Delta\Omega,
\]
with the detection probability \(\varepsilon_d\) given in
Table~\ref{tab:prob-reg}, assumed target thickness
\(t = 7.5\times 10^{20}~\text{cm}^{-2}\), the \isotope[8]{He} beam intensity
\(I = 2\times 10^5~\text{s}^{-1}\), and with the estimated values of
the solid angle covered by the triggering array 2a shown
in Figure~\ref{fig:setup}.

Data reported in \cite{Nikolskii:2022} allow for rough estimations made on
the cross sections of the
\(\isotope[2]{H}(\isotope[8]{He},\isotope[3]{He})\isotope[7]{H}\) reaction\
resulting in the population of the \isotope[7]{H} ground state and its first
excited state having energy \(approx 2.2\) MeV and \(approx 5.5\) MeV,
respectively, above the \(\isotope[3]{H} +4n decay\) threshold.
Average differential cross section estimated for the
\(\isotope[2]{H}(\isotope[8]{He},\isotope[3]{He})\isotope[7]{H}\) reaction
populating the \isotope[7]{H} ground state in angular range
\(\theta_{\text{cm}}=0^\circ–20^\circ\) is
\(1\times 10^{-29}~\text{cm}^2/\text{sr}\).
Such estimate made for the first excited state population is
\(3\times 10^{-29}~\text{cm}^2/\text{sr}\).
Consequently, total statistics gained during the one-month time length of
data taking done with the \isotope[8]{He} beam intensity
\(2 \times 10^5~\text{s}^{-1}\) on the target will contain ~50
five-body decay events recorded for each of these two \isotope[7]{H} states.

\section{Cross-talk and pile-up caused losses}

The Neutron Wall composition made with the 75-mm thick plastics results that
less than 5\% of the total number of detected 3-fold and 4-fold neutron events
are spoiled by the occurrence of such events when a single neutron is detected,
due to re-scattering, in two adjacent scintillators (the so-called cross-talk).
Time difference \(< 1\)  ns between the two neutron signals distinguishes these
cross-talks from the overwhelming part of events obtained with two
\isotope[7]{H} decay neutrons detected in adjacent scintillators.
Throwing out all events looking as cross-talks will not result in any noticeable
loss in the luminosity of the experiment.
The loss occurring from this ejection in the real collected data will be less
than 2\%.

Probability to have two neutrons hitting a single plastic makes about 8\% when
four neutrons emitted at the \isotope[7]{H} decay arrive at the Neutron Wall.
More than one-half of such pile-up events will be stored with the complete data
set, characterizing the \isotope[3]{He} recoil, and with four
(3 neutrons and \isotope[3]{H}) of the five \isotope[7]{H} decay products.
About 4\% of pile-up events are possible when three neutrons, emitted at
the \isotope[7]{H} decay, arrive at the Neutron Wall.
Therefore, conclusion is made that the neutron pile-up results in a 4\%
reduction of detection efficiency as compared with the data given in the
last row of Table~\ref{tab:prob-reg}.

In summary, taking into account the cross-talk rejection and the losses caused
due to pile-ups one should take luminosity estimates reduced by 6\% as compared
with the numbers presented in Table~\ref{tab:lum-est}.

In addition, we expect a small background contribution from gamma rays.
The main gamma sources could be materials 1, 2a, 2b and 3
(Figure~\ref{fig:setup}).
In this case, the time of flight to the Neutron Wall will be less than
\(\sim 9\) ns, which is far from the typical time of flights corresponded to
neutrons, see Figure~\ref{fig:n-spectra}.

\section{Experimental test and conclusions}

The given above estimations of cross-talk and detection efficiency,
made for the Neutron Wall looking as a very compact assembly of BC-404 plastics,
should be checked experimentally in a wide energy range of neutron energy
15--45 MeV.
For this task the measurements of the
\(\isotope[1]{H}(\isotope[3]{H},\isotope[3]{He})n\) reaction,
induced by the \isotope[3]{H} beam obtained at ACCULINNA-2 with energy
25--60~MeV and intensity \(10^6~\text{s}^{-1}\), are foreseen.
The recoil \isotope[3]{He} nuclei are emitted from the target showing
the neutron energy and flight direction.
In the \(\isotope[1]{H}(\isotope[3]{H},\isotope[3]{He})n\) reaction
cross section making \(\sim 1\) mb/sr and at the thickness of hydrogen target
\(\sim 10^{21}~\text{cm}^{-2}\) about 1 neutron per minute will be detected by
a single BC-404 detector module included into an array of 7 plastics shown in
Figure~\ref{fig:plastic}.
In these measurements the triggering process (detection of low energy,
\(\sim 10-20\) MeV, \isotope[3]{He} recoil nuclei) using Si annular telescope
will be optimized as well.

The proposed detector systems\footnote{
  Present status of the new detector array is as follows:
  1) there are all components of NW, i.~e. 100 BC-404 plastics and ETE-9822B
  photomultipliers;
  2) the annular Si detectors with 32/125 mm inner/outer diameter segmented
  in 64 rings and 64 sectors will be ready in fall 2023;
  3) the array of ten 1-meter long plastic scintillators, 88 mm wide and
  20 mm thick, is under design.
},
together with the existing array of stilbene modules~\cite{Bezbakh:2018},
will increase significantly the luminosity of the ACCULINNA-2 setup which plays
a key role in the experiments with radioactive beams~\cite{Fomichev:2018}.
These simulations, carried out for the
\(\isotope[2]{H}(\isotope[8]{He},\isotope[3]{He})\isotope[7]{H}\) reaction,
are the first approximation to the detail studies made on
the multi-neutron decay of exotic nuclei like \isotope[7]{H},
\(\isotope[7]{He}^*\) and \isotope[6]{H}, produced with the \isotope[8]{He} beam
bombarding deuterium target, and on the \(4n\) excitation spectrum obtained
in the \(\isotope[2]{H}(\isotope[8]{He},\isotope[6]{Li})4n\) reaction.
\section*{Acknowledgments}
This work was done in the frame of collaboration with National Center of
Physics and Mathematics
(project 8 ``Physics of hydrogen isotopes'', topic 8.3
``Study of properties of neutron-rich nuclei away from the stability valley'').
The activity was partly supported by the Russian Science Foundation Grant
No. 22-12-00054.
We acknowledge Prof. M.~S. Golovkov for critical remarks and recommendations.

\bibliographystyle{pepan}
\bibliography{multi-n}

\end{document}